\documentclass[letterpaper]{article} %
\usepackage{aaai2027}          %
\usepackage[hyphens]{url}
\usepackage{graphicx}
\usepackage{natbib}
\usepackage{caption}
\usepackage{amsmath}
\usepackage{booktabs}
\usepackage{tikz}
\usetikzlibrary{positioning, calc, arrows.meta}
\usepackage[capitalise,noabbrev]{cleveref}

\newcommand{\gapmissing}{\texttt{missing}}
\newcommand{\gapstale}{\texttt{stale}}
\newcommand{\gapcontradictory}{\texttt{contradictory}}
\newcommand{\gapambiguous}{\texttt{ambiguous}}
\newcommand{\gapbrokenxref}{\texttt{broken\_xref}}

\newcommand{\metricSpanDays}{43}
\newcommand{\metricTickets}{499}
\newcommand{\metricGaps}{92}
\newcommand{\metricLatencyHours}{4.6}

\newcommand{\metricBTickets}{192}
\newcommand{\metricBAssertions}{1121}
\newcommand{\metricBCorrect}{245}
\newcommand{\metricBIncorrect}{57}
\newcommand{\metricBUnverifiable}{813}
\newcommand{\metricBVerifiableAccPct}{81.1}
\newcommand{\metricBUnverifiablePct}{72.5}
\newcommand{\metricBKappa}{0.73}
\newcommand{\metricBTie}{6}   %
\newcommand{\metricBErrFloorPct}{5.1}   %

\newcommand{\metricBAccCIlo}{76.3}   %
\newcommand{\metricBAccCIhi}{85.1}
\newcommand{\metricCCompCIlo}{61.5}  %
\newcommand{\metricCCompCIhi}{78.2}
\newcommand{\metricDCIlo}{54.9}      %
\newcommand{\metricDCIhi}{83.7}

\newcommand{\metricBKappaVer}{0.50}       %
\newcommand{\metricBUnanCorrect}{79}      %
\newcommand{\metricBUnanIncorrect}{39}
\newcommand{\metricBUnanUnver}{87}

\newcommand{\metricCPairs}{192}
\newcommand{\metricCAgentPref}{150}
\newcommand{\metricCHumanPref}{35}
\newcommand{\metricCKappa}{0.41}
\newcommand{\metricCCompN}{112}            %
\newcommand{\metricCCompAgentPct}{70.5}
\newcommand{\metricCCompHumanPct}{23.2}
\newcommand{\metricCSplitPct}{3.6}         %
\newcommand{\metricCCompSplitPct}{6.2}     %
\newcommand{\metricCAgentAddsPct}{91.1}    %
\newcommand{\metricCHumanAddsPct}{83.9}    %

\newcommand{\metricDN}{35}
\newcommand{\metricDAfterPrefPct}{71.4}
\newcommand{\metricDBeforePrefPct}{28.6}
\newcommand{\metricDKappa}{0.61}
\newcommand{\metricDLoopClosureGaps}{7}    %
\newcommand{\metricDMatched}{231}
\newcommand{\metricDNotHighConf}{40}
\newcommand{\metricDNoBefore}{5}
\newcommand{\metricDNoConf}{1}

\newcommand{\metricEPairs}{192}
\newcommand{\metricEAgentPct}{64.6}        %
\newcommand{\metricEBaselinePct}{31.2}     %
\newcommand{\metricEKappa}{0.55}

\newcommand{\metricAuditNCorrect}{51}
\newcommand{\metricAuditNUseful}{30}      %
\newcommand{\metricAuditNImprove}{35}
\newcommand{\metricAuditKappaCorrect}{0.30}
\newcommand{\metricAuditKappaImprove}{0.68}
\newcommand{\metricAuditAgrImprovePct}{83}
\newcommand{\metricDAfterPrefHumanPct}{74.3}   %

\newcommand{\metricProbeBCorrect}{279}      %
\newcommand{\metricProbeBIncorrect}{164}    %
\newcommand{\metricProbeBUnverifiable}{678} %
\newcommand{\metricProbeBLeanCorrect}{63}   %
\newcommand{\metricProbeBLeanPanel}{29}     %
\newcommand{\metricProbeCLeanAgent}{4}      %
\newcommand{\metricProbeCLeanPanel}{48}     %
\newcommand{\metricProbeDAfter}{30}         %
\newcommand{\metricDAfterN}{25}             %
\newcommand{\metricRegistryTargets}{21}     %
\newcommand{\metricWatchedPackages}{18}     %

\title{Mind the Gap: A Deployed Multi-Agent System for Documentation Maintenance}

\author{
    Lorenzo Masoero
}
\affiliations{
    Amazon\\
    masoerl@amazon.com
}

\begin{document}
\maketitle

\begin{abstract}
The ``AI revolution'' has enabled teams across the technology industry to deliver software features at unprecedented velocity.
While code evolves faster than ever, customer-facing documentation often struggles to stay up to speed.
As users increasingly rely on LLM-based agents that consume documentation as their primary interface, the impact of this documentation gap worsens: stale information propagates and is treated as authoritative.
In this paper, we argue that just as AI has automated code delivery, it must also automate the maintenance of the documentation that describes it.
Signals for identifying where documentation has fallen behind already exist in abundance---both explicit (e.g., resolved support tickets revealing what documentation lacks) and implicit (e.g., code-documentation discrepancies detectable from automated loops).
We deploy a multi-agent system, built around a triage agent we call \emph{``TriAgent''}, that continuously extracts these signals and repairs documentation drift.
Over \metricSpanDays{}~days of shadow-mode operation, our system triaged \metricTickets{}~tickets and surfaced \metricGaps{}~gaps in the documentation.
Using a panel of LLM judges from independent model families, \metricBVerifiableAccPct{}\% of TriAgent's \emph{verifiable} assertions were correct (\metricBCorrect{} of the 302 assertions that the human resolution addresses, out of \metricBAssertions{} extracted; majority of three cross-family judges); on length-comparable pairs the panel preferred TriAgent's response in \metricCCompAgentPct{}\% of pairs and the human engineer's in \metricCCompHumanPct{}\% (the rest had no judge majority). Importantly, agent- and human-sourced replies tended to be complementary, with the human often surfacing insights that were either buried within the existing documentation or missing from it entirely; in turn, our system turns each such insight into an explicit gap and, after human review, updates the customer-facing documentation that users and their LLM agents consume.
Repairing a gap measurably improved the agent's later answer: when the agent re-answered the same ticket after the repair, blinded judges preferred the post-repair answer over its own pre-repair answer \metricDAfterPrefPct{}\% of the time ($n=\metricDN{}$), and the documentation commit that resolved each tagged gap postdated its discovery. A blinded human audit confirms strong alignment on self-improvement.
\end{abstract}

\section{Introduction}
\label{sec:intro}

Most teams in large technology companies have adopted AI-driven development practices that dramatically accelerate the rate at which the codebases powering the services they own change \citep{peng2023copilot}: 
features that previously required weeks of engineering effort now reach production in days.
Ensuring that code obtained via AI-driven development meets the same standards as human development is a well-understood priority. 
Yet as code delivery accelerates, customer-facing documentation often risks falling behind. And, absent an explicit mechanism, this gap is likely to be left unaddressed and grow over time.

Risks related to stale documentation compound, as LLMs are ubiquitous across code development and in how users consume services: documentation is the primary entry point for both human users and their LLM-based tools.
On paper, this creates a strong incentive for service teams to maintain accurate, exhaustive documentation, and to offer a specialized support agent constrained to authoritative sources they control.
When documentation is exhaustive, such agents are more likely to provide accurate answers to service inquiries than a custom agent executed by a customer of the service, precisely because these agents can be built so as to be constrained to authoritative, curated sources.
Yet this opportunity is, in our experience, widely under-appreciated in practice: documentation is maintained reactively, in isolation, and without standard practices.
In this paper, we argue that even when the in-house agent cannot answer accurately, the attempt is still worthwhile: the failure itself is a signal that reveals where documentation falls short, and a system built to extract and act on that signal can continuously close the gap.
To explain the problem, and our solution, we consider three customer contacts of escalating difficulty in the context of a production support queue for a large A/B testing service.

\paragraph{Simple scenario: Routine questions, sufficient documentation.}
The first case is a customer contact with a routine question.
A user reports that their experiment shows zero traffic before a certain date and asks whether this is a bug.
An agent consulting the service's documentation identifies the cause immediately: the experiment was configured with a gradual ramp schedule.
The date in question is when traffic allocation began, not when the experiment was created.
This answer, grounded entirely in existing documentation, is identical to what the human responder provides.
The documentation is sufficient. The system effectively replaces the oncall duty a human engineer would have otherwise attended, automating and streamlining the process. 

\paragraph{Harder scenario: Visible gaps from human corrections.}
In a second question from the same queue, a user reports that a long-running analysis job disappeared from the portal.
The agent, drawing on the same documentation, suggests plausible explanations: the job may have been superseded by an automated pipeline, or it may be hidden in the default view.
The human responder, however, identifies the actual cause: a scheduled infrastructure patch reset the computation host mid-job, and the restarted job faces a runtime limit not documented anywhere.
The agent's answer is coherent and grounded, but wrong.
Comparing the agentic and human responses directly reveals what the documentation lacks, without requiring any manual annotation or audit.
This case is addressable with human-in-the-loop supervision at the point of customer contact: the human corrects the agent's response, and the divergence signals a documentation gap.
The system does not perform root-cause analysis: it simply extracts the output of the human reasoning. 
Once documented, the agent can handle the same symptom correctly.

\paragraph{Hardest scenario: Invisible drift between code and documentation.}
The hardest case for our setting is one where a code change alters system behavior, but no user happens to ask about it and no human corrects an agent.
In this case, the documentation quietly falls out of sync with production.
This latent drift accumulates until it surfaces as a customer-facing failure, or worse, as a quietly incorrect agent response that no one challenges.
It is only addressable if an end-to-end system tracks code development and proactively surfaces documentation that has fallen behind.

Motivated by these three examples, we argue that the same standards and continuous pipelines that govern code delivery must extend to the documentation that describes it.
To fill this gap, we deploy a multi-agent system that automates support responses and continuously repairs the documentation gaps they expose, both visible gaps revealed by human corrections and latent gaps where code outpaces documentation.
Over \metricSpanDays{}~days of production (shadow-mode) deployment on the service's support queue, our system triaged \metricTickets{}~tickets and discovered \metricGaps{}~documentation gaps.
We employed LLM-as-a-judge evaluation of the agentic replies, using three judges from independent model families (see \cref{sec:eval} for more details).
We extracted \metricBAssertions{}~factual assertions from \metricBTickets{}~agent responses. The human resolution addresses 302 of them, and \metricBCorrect{} of those (\metricBVerifiableAccPct{}\%) were confirmed correct. Most of the remaining assertions are not addressed by the human reply and cannot be checked against it. The errors that remain predominantly involve operational specifics absent from documentation.
On length-comparable ticket pairs the LLM judges preferred the agent's documentation-grounded response over the human in \metricCCompAgentPct{}\% of cases. 
Human responses were preferred in \metricCCompHumanPct{}\% of cases (the rest had no judge majority).
Importantly for our system, agent and human each raised correct points the other missed; where the human is preferred, their operational additions surface actionable gap signals that feed back into the maintenance loop.
Repairing a gap, in turn, measurably improved the agent's later answer: when the agent re-answered the same ticket after the repair, the post-repair answer was preferred over its own pre-repair answer in \metricDAfterPrefPct{}\% of blinded comparisons (\metricDAfterPrefHumanPct{}\% in a blinded human audit of the same \metricAuditNImprove{}~pairs).

\paragraph{Contributions.}
We make three contributions.
First, we reframe documentation maintenance as an inference problem: resolved support tickets are implicit supervision. The discrepancy between what the documentation provides and what a human responder supplies is itself the gap signal, and it costs nothing to collect.
Second, we build and deploy a multi-agent system that acts on this signal, detecting gaps from human corrections, proactive probing, and code--documentation drift, and drafting fixes that a human reviews before merge.
Third, we report a \metricSpanDays{}-day production deployment---measuring factual accuracy and agent--human agreement, and distilling the operational lessons of running the loop, including where it is throttled in practice.

\section{System}
\label{sec:system}

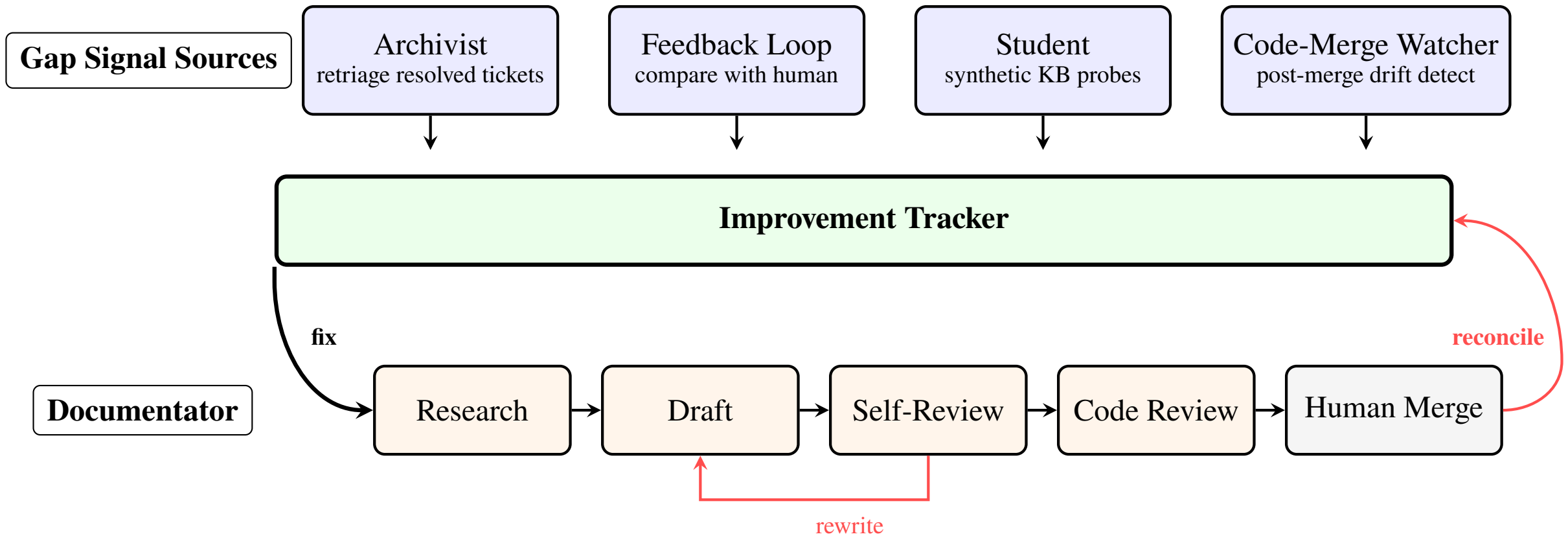
\begin{figure*}[t]
\centering

\begin{tikzpicture}[
    node distance=0.4cm and 0.6cm,
    every node/.style={font=\small},
    src/.style={draw, rounded corners=3pt, fill=blue!8, minimum height=1.1cm, minimum width=2.6cm, align=center, thick},
    staged/.style={draw, rounded corners=3pt, fill=gray!8, minimum height=1.1cm, minimum width=2.6cm, align=center, thick, dashed},
    hub/.style={draw, rounded corners=3pt, fill=green!8, minimum height=0.9cm, minimum width=12cm, align=center, very thick},
    pipe/.style={draw, rounded corners=3pt, fill=orange!8, minimum height=0.9cm, minimum width=2cm, align=center, thick},
    outp/.style={draw, rounded corners=3pt, fill=gray!8, minimum height=0.9cm, minimum width=2.2cm, align=center, thick},
    arr/.style={->, >=stealth, thick},
    darr/.style={->, >=stealth, thick, dashed},
    lbl/.style={font=\scriptsize, text=gray},
]

\node[src] (harvest) {Archivist\\[-2pt]{\scriptsize retriage resolved tickets}};
\node[src, right=0.5cm of harvest] (feedback) {Feedback Loop\\[-2pt]{\scriptsize compare with human}};
\node[src, right=0.5cm of feedback] (student) {Student\\[-2pt]{\scriptsize synthetic KB probes}};
\node[src, right=0.5cm of student] (watcher) {Code-Merge Watcher\\[-2pt]{\scriptsize post-merge drift detect}};

\node[draw, rounded corners=2pt, fill=white, inner sep=4pt, font=\small\bfseries, left=0.1cm of harvest.west, anchor=east] (src_title) {Gap Signal Sources};

\draw[arr] (harvest.south) -- ++(0,-0.35);
\draw[arr] (feedback.south) -- ++(0,-0.35);
\draw[arr] (student.south) -- ++(0,-0.35);
\draw[arr] (watcher.south) -- ++(0,-0.35);

\node[hub, below=0.6cm of feedback.south, xshift=1.3cm] (tracker) {\textbf{Improvement Tracker}};

\node[pipe, below=1.0cm of tracker.south, xshift=-4cm] (research) {Research};

\node[draw, rounded corners=2pt, fill=white, inner sep=4pt, font=\small\bfseries, anchor=east] (doc_title) at ([xshift=-0.4cm]src_title.east |- research) {Documentator};
\node[pipe, right=0.3cm of research] (draft) {Draft};
\node[pipe, right=0.3cm of draft] (review) {Self-Review};
\node[pipe, right=0.3cm of review] (cr) {Code Review};
\node[outp, right=0.3cm of cr] (merge) {Human Merge};

\draw[arr, very thick] (tracker.south west) -- ++(0,-0.15) to[out=-90, in=180] (research.west);
\node[font=\scriptsize\bfseries] at ([xshift=-0.5cm, yshift=0.75cm]research.west) {fix};
\draw[arr, red!70, thick] (merge.east) to[out=0, in=-90] ++(0.6,0.5) to[out=90, in=0] (tracker.east);
\node[font=\scriptsize\bfseries, text=red!70] at ([xshift=-0.05cm, yshift=0.75cm]merge.east) {reconcile};

\draw[arr] (research) -- (draft);
\draw[arr] (draft) -- (review);
\draw[arr] (review) -- (cr);
\draw[arr] (cr) -- (merge);

\draw[arr, red!70] (review.south) -- ++(0,-0.45) -| (draft.south);
\node[lbl, text=red!70, below=0.5cm of review.south, xshift=-0.8cm] {rewrite};

\end{tikzpicture}

\caption{System architecture. Gap signal sources (top) feed the improvement tracker through deduplication. The Documentator (bottom) drains the tracker through a fixed pipeline with a blocking self-review gate. Red arrows show feedback loops: rewrite on self-review failure, and reconciliation from human merge back to the tracker.}
\label{fig:architecture}
\end{figure*}

Motivated by the three scenarios of \cref{sec:intro}, we propose a system that automates support responses and turns two signals into documentation fixes: whether each answer succeeds or fails (reactive), and code--documentation drift caught before any ticket is filed (proactive).
We organize it around two problems: detecting where documentation fails (\cref{sec:detecting}) and producing correct fixes (\cref{sec:fixing}).
Five agents coordinate through a shared backlog, the improvement tracker, following a multi-agent conversation pattern \citep{wu2024autogen}.
\Cref{fig:architecture} shows the full signal flow.

\begin{table*}[t]
\centering
\small
\begin{tabular}{lllp{7.5cm}}
\toprule
\textbf{Agent} & \textbf{Stage} & \textbf{Mode} & \textbf{Role} \\
\midrule
TriAgent & Triage & Live & First-response classification/drafting for incoming tickets \\
Feedback Loop & Detection & Live & Blind re-triage of resolved tickets, structured learning \\
Archivist & Detection & Batch & Gap discovery (harvest) and remediation orchestrator \\
Student & Detection & Proactive & Adversarial and speculative probing of KB topics \\
Code-Merge Watcher & Detection & Proactive & Post-merge documentation-drift detection \\
Documentator & Remediation & Batch & Research $\rightarrow$ draft $\rightarrow$ self-review $\rightarrow$ code review \\
\bottomrule
\end{tabular}
\caption{System agents. \emph{Stage} indicates the agent's role in the loop (triage, detection, or remediation). \emph{Mode} indicates when it runs. Shared state is the integration layer; no agent calls another directly.}
\label{tab:agents}
\end{table*}

\paragraph{One ticket, end to end.}
To make the flow concrete, consider the disappeared-analysis-job ticket from \cref{sec:intro}.
``\emph{TriAgent}'' drafts a documentation-grounded response and surfaces it to the on-call responder, who instead supplies the true cause: a host reset mid-job and an undocumented runtime limit.
The Feedback Loop agent compares the two responses, isolates the missing constraint, and writes a gap entry to the improvement tracker.
The Documentator later drains that entry---researching the target document, drafting a fix, passing it through self-review, and opening a code review that a human approves before merge.
Once merged, the same symptom is answerable from documentation alone.

\subsection{Detecting Gaps}
\label{sec:detecting}

The first stage of the system revolves around learning about existing documentation gaps.
As per \cref{sec:intro}, these gaps manifest in two modes: visibly, through support tickets that signal something is wrong; and invisibly, through silent divergence that no user reports.
We address both.

\paragraph{Visible drift.}
Sometimes, when the documentation describing the system is unclear or stale, users file ``tickets'' in a dedicated support queue (i.e., complaints or questions about the service).
In this case, the support queue can serve as a detection mechanism for the system.
To convert this signal into actionable gap detection, the system operates a live triage loop via TriAgent: when a ticket arrives, it drafts a proposed response using only documentation-grounded evidence.
This proposed response is surfaced to the on-call human who is responsible for resolving the ticket.
The human is de facto acting as a specialized judge.
If they approve the agent's response -- which is grounded exclusively in authoritative documentation -- this provides evidence that existing documentation is sufficient for that class of question.
If instead the human provides a different response, this is evidence of a gap in authoritative documentation.
We close this gap by activating a ``\emph{Feedback Loop}'' agent, which compares the two responses, identifies the missing knowledge, and writes a structured gap entry to the improvement tracker.
Both outcomes are informative: approval confirms coverage; divergence reveals gaps.

This same mechanism extends backwards in time.
A ``\emph{Harvest}'' mode, orchestrated by the ``\emph{Archivist}'' agent, applies the identical blind-comparison procedure to tickets already resolved in the past, backfilling documentation gaps from historical support data without requiring any new human effort.

\paragraph{Invisible drift.}
Not all documentation failures generate tickets, particularly when LLM-based consumers form incorrect beliefs rather than escalating.
We address this through two mechanisms.
First, via a ``\emph{Student}'' agent, whose goal is to stress-test documentation coverage by generating synthetic probes: questions on topics not yet seen in the support queue, surfacing coverage gaps before a user encounters them.
Second, a ``\emph{Code-Merge Watcher}'', a dedicated agent whose goal is to detect documentation drift at the source, monitors production code packages for merged changes, performs an inverse lookup from changed code to affected documentation through a code-to-documentation registry (\cref{app:registry}), and flags cases where documentation may have drifted as a result of a code change, catching gaps at merge time, before any user is affected.
This represents a shift from reactive gap detection to integration with the development process itself: rather than waiting for drift to surface through support tickets, the system intercepts documentation-affecting code changes at merge time and can initiate fixes before the change reaches users.
Gaps discovered through either mechanism are written to the same tracker as ticket-derived gaps, with no distinction in how they are prioritized or resolved.

\subsection{Producing Fixes}
\label{sec:fixing}

The next logical step, after having surfaced gaps as described in \cref{sec:detecting}, is to resolve them.
All detection signals converge on the improvement tracker, a shared backlog that deduplicates and prioritizes gaps automatically.

We resolve gaps via a dedicated ``\emph{Documentator}'' agent that drains the improvement tracker, converting gap signals into reviewed documentation updates through a fixed pipeline.
This agent researches the target document, and drafts a fix matching the target format.
A natural concern is that AI-generated fixes may themselves contain errors.
We address this risk in two ways.
First, before any fix reaches a human reviewer, we gate it through a structured ``\emph{self-review}'' that evaluates the draft against factual consistency and scientific rigor criteria.
Second, a human is always in the loop: every fix produced by the Documentator is submitted as a code review that requires explicit human approval before merging.
The human's role shifts from author to reviewer -- but the review authority remains with a person.

\paragraph{Data governance.}
The system operates on internal support tickets that can contain customer identifiers and production details.
All processing runs inside the service team's existing access-controlled environment, and the agents inherit the same authorization scope as the on-call engineers they assist; ticket content is not exported outside that boundary.
The examples and evaluation artifacts reported in this paper were reviewed to remove customer-identifying and service-confidential specifics.

\section{Evaluation}
\label{sec:eval}

We evaluate the deployed system in five parts: its operational scale, the factual correctness of its responses, their usefulness relative to human resolutions, whether repairing a gap measurably improves later answers, and whether the full loop beats plain retrieval. Results are drawn from a frozen \metricSpanDays{}-day pilot on a production support queue for a causal-inference (A/B testing) analysis service.

All automated judgments use a panel of three judges drawn from \emph{different model families than the agent's generation model} (Model~A); we report the majority verdict and inter-judge agreement (Fleiss' $\kappa$). We exclude same-family judges from the headline because prior work finds that a judge tends to favor outputs from its own family \citep{spiliopoulou2025playfavorites}. A fourth, same-family judge is run separately as a probe (\cref{app:judges}). The \metricTickets{}~triaged tickets measure deployment scale and are a distinct population from the scored set. Quality is graded on the \metricCPairs{}~pairs that received both an agent and a human response, after excluding purely operational requests (e.g., ``clear caches'') and trivially short exchanges and dropping a handful of non-answers the agent correctly declined (full funnel in \cref{app:dataset}).

\subsection{Deployment Metrics}
\label{sec:operational}

\begin{table}[t]
\centering
\small
\begin{tabular}{lr}
\toprule
\textbf{Metric} & \textbf{Value} \\
\midrule
Tickets triaged & \metricTickets{} \\
Documentation gaps discovered & \metricGaps{} \\
Median agent response availability lead time & \metricLatencyHours{}\,h \\
Agent drafting time (gap $\to$ submitted fix) & $<$5\,min \\
\bottomrule
\end{tabular}
\caption{Operational scale over the \metricSpanDays{}-day pilot. \emph{Response availability lead time} is the median time from ticket arrival to an agent-drafted answer being available.}
\label{tab:operational}
\end{table}

Table~\ref{tab:operational} summarizes the pilot. The system triaged \metricTickets{}~tickets and surfaced \metricGaps{}~documentation gaps (per-mechanism breakdown in \cref{app:taxonomy}). The proactive Code-Merge Watcher recorded no gaps in this window: it fires only at code-merge time and its reach is bounded by the code-to-documentation registry (see Limitations). The agent drafts a candidate fix in under five minutes. A human reviews every draft before it merges. The tracker marks all \metricGaps{}~gaps as fixed, most of them in a single batch at the end of the window; \metricDLoopClosureGaps{} carry a gap-tagged commit in the documentation history.

\begin{figure}[h!]
\centering
\includegraphics[width=.7\columnwidth]{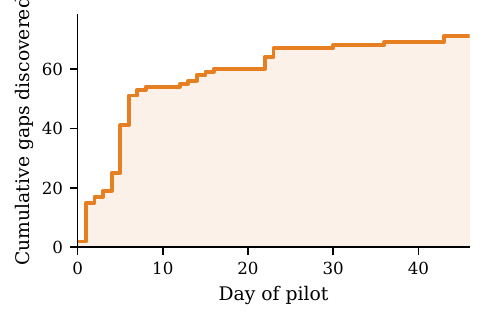}
\caption{Documentation gaps over the \metricSpanDays{}-day pilot, discovered across multiple mechanisms during the window.}
\label{fig:cumulative-gaps}
\end{figure}

\subsection{Factual Correctness}
\label{sec:factual}

We evaluate factual correctness at the assertion level with an automated pipeline that sits outside the deployed system, simply grading (and not producing) the agent's answers. An LLM extraction pass identifies the factual claims in each TriAgent response; three LLM judges then label every claim independently against the human responder's resolution (the single authoritative reference). We report their majority label.
We adopt a mutually-exclusive rule: a claim made by the TriAgent is \textit{incorrect} if the human reference contradicts it, \textit{correct} if the human reference addresses and supports it, and \textit{unverifiable} if the human reference does not speak to it. 

From \metricBTickets{}~responses we extracted \metricBAssertions{}~assertions. 
By majority the panel labels \metricBCorrect{}~correct, \metricBIncorrect{}~incorrect, and \metricBUnverifiable{}~(\metricBUnverifiablePct{}\%) unverifiable.
On average, judges agree closely with one another (Fleiss' $\kappa=\metricBKappa{}$, ``substantial''), i.e.\ the verdicts do not hinge on any single model.
On \metricBTie{}~of the \metricBAssertions{}~assertions the three judges split three ways (one correct, one incorrect, one unverifiable). These assertions are discarded for our evaluation. 
We define \emph{accuracy} as the fraction of assertions on which the TriAgent and the human agree. Because the human's resolution is usually brief and answers only what was asked, much of what the agent adds is not mentioned in the human reply; we label these claims \emph{unverifiable}, since the reference neither confirms nor contradicts them. We therefore report accuracy over the \emph{verifiable} assertions, where it is \metricBVerifiableAccPct{}\% (\metricBCorrect{}/302; Wilson 95\% CI [\metricBAccCIlo{}, \metricBAccCIhi{}], see \cref{app:ci}). If instead we count every unverifiable claim as an error, accuracy drops to 21.9\%, though most such claims are probably fine. The firm number is the error floor: \metricBIncorrect{} of \metricBAssertions{} assertions (\metricBErrFloorPct{}\%) directly contradicted the human.
The same-family probe is more decisive than the panel in both directions: it tags more assertions \textit{correct} (\metricProbeBCorrect{} vs.\ \metricBCorrect{}) and more \textit{incorrect} (\metricProbeBIncorrect{} vs.\ \metricBIncorrect{}; \cref{app:judges}).

Each of the \metricBIncorrect{}~contradicted assertions points to a specific place where the agent's sources were wrong or incomplete, and each feeds back into the gap-detection loop. Validating the \emph{categorization} of these errors against human labels is deferred to a human-audit round (see Limitations).

\subsection{Usefulness}
\label{sec:pairwise}

\begin{table}[t]
\centering
\small
\begin{tabular}{lcc}
\toprule
\textbf{Outcome} & \textbf{All pairs} & \textbf{Comparable length} \\
\midrule
Agent preferred & 78.1\% & \metricCCompAgentPct{}\% \\
Human preferred & 18.2\% & \metricCCompHumanPct{}\% \\
No majority & \metricCSplitPct{}\% & \metricCCompSplitPct{}\% \\
\bottomrule
\end{tabular}
\caption{Blinded pairwise preference (\metricCPairs{}~pairs, majority of three cross-family judges, $\kappa=\metricCKappa{}$). The comparable-length column restricts to the \metricCCompN{}~pairs whose two responses are within a $2\times$ length ratio.}
\label{tab:pairwise}
\end{table}

A correct answer is valuable only if it is also \emph{useful}---one that actually resolves the customer's ticket, at least as well as a human expert would. We measure this directly with blinded pairwise preference, scored by the same three-judge cross-family LLM panel: for each ticket, each judge sees the agent's response and the human's (order-counterbalanced, with the source hidden) and picks which one better resolves it, and we report the majority. Across all \metricCPairs{}~pairs the agent is preferred on \metricCAgentPref{} (78.1\%) and the human on \metricCHumanPref{} (18.2\%). This raw figure is confounded by length: where the agent's response is more than twice as long it is preferred in every case, and where it is much shorter the human wins. Restricting to length-comparable pairs (Table~\ref{tab:pairwise}) the agent is still preferred \metricCCompAgentPct{}\% to \metricCCompHumanPct{}\% (95\% CI on the former [\metricCCompCIlo{}, \metricCCompCIhi{}])---so the preference is substantial but partly a verbosity effect.

Beyond who is preferred, the agent and human responses are largely complementary: judged on coverage rather than winner, the agent raises a correct, material point the human omits in \metricCAgentAddsPct{}\% of pairs, and the human raises one the agent omits in \metricCHumanAddsPct{}\%. The agent broadens coverage, while the human's additions are typically operational knowledge---permissions, escalations, transient system state---absent from documentation. Each such human-added point is a documentation defect the system can localize and repair; the preference and coverage signals feed directly into the gap-detection loop of \cref{sec:detecting}.

\subsection{Self-Improvement}
\label{sec:selfimprove}

We next test the system's central claim: that fixing the documentation improves the quality of the replies the agent later produces.
For the \metricDN{}~gaps where the agent's original response was not high-confidence (its self-reported \textsc{medium}/\textsc{low} label)---the cases where improvement is a meaningful question---we compare that original answer with a later re-answer produced against the updated documentation. Both answers use the same agent (Model~A), prompt, and ticket, with no access to the human resolution; only the retrieved documentation differs, isolating the fix as the cause of any improvement. Scoring is blinded: the judge sees the two answers without being told which came first, and reports which better resolves the ticket. 
The later answer is preferred in \metricDAfterPrefPct{}\% of cases and the earlier in \metricDBeforePrefPct{}\% ($\kappa=\metricDKappa{}$; 95\% CI on the \metricDAfterPrefPct{}\%: [\metricDCIlo{}, \metricDCIhi{}], $n=\metricDN{}$; corroborated by a blinded human audit, see \cref{sec:limitations}).

A skeptic might object that the documentation was there all along, so the fix changed nothing. We rule this out via the edit history: for every tagged gap, the commit updating the relevant page postdates our flagging it, so the content that improved the second answer did not exist when the first was written.

\subsection{Is It More Than Retrieval?}
\label{sec:baseline}

Finally, does the full agent pipeline earn its complexity over plain retrieve-and-generate? We build a retrieval-only baseline---BM25, a standard lexical ranking function \citep{robertson2009bm25}, over the same documentation corpus, feeding the top passages to the \emph{same} generation model as the agent (Model~A), with no gap detection, feedback, or self-improvement---and run the blinded pairwise comparison of \cref{sec:pairwise} between the full agent's deployed response and the baseline's over the \metricEPairs{}~pairs. The full system is preferred \metricEAgentPct{}\% of the time and the baseline \metricEBaselinePct{}\% ($\kappa=\metricEKappa{}$). Both systems retrieve over the same documentation, which the loop had already improved. They differ in retrieval and orchestration. This comparison therefore measures the full answering pipeline against plain retrieval. The before/after comparison of \cref{sec:selfimprove} measures the maintenance loop itself, holding the pipeline fixed and changing only the documentation.

\section{Related Work}
\label{sec:related}

Deploying large language models for operational support is increasingly common, from root cause analysis of cloud incidents \citep{chen2024rca,zhang2024icl_rca} to autonomous multi-step diagnosis \citep{wang2024rcagent,roy2024rca_agents} and design frameworks for autonomous cloud operations \citep{shetty2024autonomous_clouds,chen2025aiopslab}.
These systems retrieve from documentation or knowledge bases via retrieval-augmented generation \citep{lewis2020rag,guu2020realm}, but uniformly treat the knowledge base as a static input. When documentation is incomplete or stale, the system degrades silently. Our work addresses this assumption: we close the loop so that operation itself repairs the knowledge base it depends on.

Self-improving agents could in principle address this gap. Reflexion \citep{shinn2023reflexion} and Self-Refine \citep{madaan2023selfrefine} demonstrate that self-critique improves task performance, and SWE-bench \citep{jimenez2024swebench} applies agents to code repair. However, these approaches rely on automated correctness oracles (test suites, execution traces) that documentation lacks. Our self-review shares the critique-then-revise structure but substitutes human review as the terminal authority.

Documentation staleness is a well-documented problem in software engineering \citep{aghajani2019documentation}, but prior automated solutions focus on code-comment synchronization or API-doc generation from source \citep{panthaplackel2020updating,mastropaolo2021studying}, where static analysis can detect drift between code and its inline annotations. Our work tackles a different problem: maintaining human-authored knowledge bases describing system behavior, operational procedures, and troubleshooting guidance. These artifacts drift from production reality in ways no static analysis can catch, because the gap is between what the system does and what prose says it does. Two concurrent 2026 systems target documentation maintenance directly: DocSync \citep{badrinarayan2026docsync} runs a critic-guided reflexion loop grounded in source-code structure, and ArtifactSync \citep{alor2026artifactsync} performs commit-level change-impact analysis. Both are \emph{code-change driven}---they detect drift by inspecting source and diffs. Our Code-Merge Watcher (\cref{sec:detecting}) works the same way. Our ticket-derived signal is orthogonal and complementary: the delta between an agent's documentation-grounded answer and the human resolution surfaces where documentation fails \emph{users}, not where it diverges from \emph{code}. A code-diff method cannot flag a document whose underlying code never changed but which users repeatedly misunderstand; ticket-derived implicit supervision can. More broadly, learning from implicit human feedback is well-studied in information retrieval \citep{joachims2005clickthrough} and language model alignment \citep{ouyang2022instructgpt}; to our knowledge, this operational-feedback signal has not previously been operationalized for documentation maintenance.

\section{Conclusion}
\label{sec:conclusion}

Documentation maintenance is usually framed as an authoring problem. We reframe it as continuous inference: resolved support tickets are implicit supervision, and the delta between what documentation says and what a human actually needed is a gap signal requiring no annotation.
We built a deployed system that exploits this signal. Over \metricSpanDays{}~days, it triaged \metricTickets{}~tickets and discovered \metricGaps{}~documentation gaps. Of the 302 agent assertions that the human resolution addresses, \metricBVerifiableAccPct{}\% were correct (Fleiss' $\kappa=\metricBKappa{}$). A blinded human audit most strongly corroborates the before/after self-improvement result ($\kappa=\metricAuditKappaImprove{}$); it endorses the offline correctness and usefulness rates only fairly, so we read those as panel-internal estimates (see Limitations).

We view our main contribution as proposing documentation maintenance as an inference problem. With this angle, resolved support tickets become a continuous, freely available supervision signal for closing documentation drift. Across \metricCPairs{}~blinded comparisons the agent's response was preferred to the human's on length-comparable pairs (\metricCCompAgentPct{}\%), each response surfaced correct points the other missed, and repairing a gap measurably improved the agent's later answer (\metricDAfterPrefPct{}\% preferred, blinded); each ticket where the human responder added value simultaneously generated an actionable gap signal, requiring no annotation, no audit, and no manual triage.

Beyond repair, the same signal informs prevention. Patterns in resolved tickets reveal which documentation structures prevent user confusion and which generate it. A team deploying a new feature can consult this signal to write effective documentation from day one, so the loop reaches from past drift to current fixes, and from past failures to future design.
While we evaluate on a single service, the underlying \emph{signal}---resolved tickets as implicit supervision---applies to any service with a documentation-backed support queue where humans correct or supplement automated responses. The specific detection heuristics, gap taxonomy, and fix pipeline would, however, require adaptation to a new domain; we do not claim the design choices transfer unchanged (see \cref{sec:limitations}).
The path to full deployment is incremental: widening from shadow-mode triage toward human-gated live responses, and from the pilot's batch merges toward rolling, per-gap repair.
See the appendices for details.

\section{Limitations}
\label{sec:limitations}

We flag five limitations of this pilot.

\textbf{Single deployment context.} We have evaluated the system exclusively on one service's support queue---a causal-inference service with a technically sophisticated user base and well-structured documentation. The architecture is domain-agnostic in principle, but services with different characteristics (higher ticket volume, less structured knowledge bases, non-technical end users, or fundamentally different failure modes) may require different detection heuristics, gap taxonomies, or fix pipelines.

\textbf{Evaluation methodology.} Correctness, preference, and before/after improvement are scored by a panel of three judges from different model families than the generation model, by majority vote with Fleiss' $\kappa$. A fourth, same-family judge is kept out of the headline as a precaution against the family favoritism reported by \citet{spiliopoulou2025playfavorites}; its verdicts differ from the panel's in both directions (\cref{app:judges}). To check the panel against human judgment, one author re-labeled a blinded, stratified subsample of the frozen decisions (\metricAuditNCorrect{} correctness assertions, \metricAuditNUseful{} usefulness pairs, \metricAuditNImprove{} before/after pairs), blind to the panel verdict and to which response was the agent's. Agreement is \emph{substantial} on the before/after judgments that carry our central self-improvement claim (Cohen's $\kappa=\metricAuditKappaImprove{}$, \metricAuditAgrImprovePct{}\% raw) and \emph{fair} on correctness and usefulness ($\kappa\approx\metricAuditKappaCorrect{}$); the annotator simply tied more readily and read correctness more leniently than the panel. We therefore treat the self-improvement result as human-validated and the offline correctness and usefulness rates as panel-internal estimates. With a single annotator these rates remain observational, and a larger multi-annotator audit is future work. The before/after set is selected: it covers the \metricDN{}~repaired gaps whose earlier answer was stored and not high-confidence, and it re-answers only the triggering ticket. Improvement on new related tickets is not yet measured. Pairwise preference is partly confounded by response length, which we mitigate by reporting the length-comparable subset alongside the full set.

\textbf{Operational boundary.} In most tickets the human responder adds operational specifics (granting permissions, clearing caches, merging code) that the agent cannot perform from documentation alone; extending coverage would require tool-use capabilities beyond KB retrieval.

\textbf{Human in the loop.} In our system, we require human review of every documentation change the agent proposes before it merges. This is a deliberate choice: the system is still maturing, and a wrong fix is costly because it becomes an authoritative answer for many downstream users and their agents. A person therefore remains the correctness and authority gate, with the agent shifting from author to reviewer. We did not measure the time reviewers spend inspecting, correcting, and merging drafts; the five-minute drafting time excludes it. Whether the loop saves human effort overall remains open.

\textbf{Proactive detection coverage.} The Code-Merge Watcher's coverage depends on the completeness of the code-to-documentation registry; unmapped packages are invisible to it. The watcher recorded no gaps in the pilot window; prevention at merge time is not yet demonstrated.

\bibliographystyle{aaai2027}
\bibliography{references}

\appendix

\section{Gap Type Taxonomy and Improvement Tracker}
\label{app:taxonomy}

The improvement tracker is the shared data structure that connects gap detection (\cref{sec:detecting}) to gap remediation (\cref{sec:fixing}). Every gap entry records a structured lifecycle: discovery metadata (source mechanism, originating ticket, timestamp), a typed classification, a priority assignment, and resolution state. We describe the classification scheme and the lifecycle below.

\subsection{Classification}

Every gap discovered by the system is classified into one of five types (Table~\ref{tab:taxonomy}): \gapmissing{} (no documentation exists), \gapstale{} (outdated behavior), \gapcontradictory{} (two sources conflict), \gapambiguous{} (underspecified), and \gapbrokenxref{} (dead or incorrect link).

The type determines the remediation path.
Gaps of type \gapmissing{} are routed to a \emph{Gap-Filler} pipeline that drafts new content from scratch using production source code as ground truth.
The remaining four types describe defects in \emph{existing} documentation and are routed to a \emph{Stale-Updater} pipeline that modifies the relevant section in place.

\paragraph{Product defects.} Some tickets reveal a defect in the product rather than in its documentation. The triage step routes every ticket by topic to the owning engineering team, which receives product defects for a code fix. TriAgent records a gap only when the documentation itself falls short, and the agents never modify product code.

\begin{table*}[h]
\centering
\small
\begin{tabular}{lp{4.5cm}p{6cm}}
\toprule
\textbf{Type} & \textbf{Definition} & \textbf{Example} \\
\midrule
\gapmissing{} & No KB coverage for the topic & A feature ships with no documentation page \\
\gapstale{} & Coverage exists but references outdated behavior & A threshold changed in code but docs still show the old value \\
\gapcontradictory{} & Two KB sources conflict & Tutorial says ``method A''; reference guide says ``method B'' \\
\gapambiguous{} & Coverage exists but is underspecified & Docs say ``may take some time'' with no latency bound \\
\gapbrokenxref{} & Link or pointer is dead or incorrect & A ``see also'' pointer targets a moved page \\
\bottomrule
\end{tabular}
\caption{Gap type taxonomy. \gapmissing{} routes to Gap-Filler; the other four route to Stale-Updater.}
\label{tab:taxonomy}
\end{table*}

In our deployment, the vast majority of gaps ($>$90\%) are of type \gapmissing{}, reflecting that documentation absence---rather than staleness---is the primary failure mode for the service studied. This finding may not generalize: services with rapidly evolving configurations may see a higher proportion of \gapstale{} gaps, while services with multiple documentation maintainers may encounter more \gapcontradictory{} cases.

\subsection{Lifecycle}

Each gap progresses through a fixed lifecycle:

\begin{enumerate}
\item \textbf{Discovery.} A detection mechanism (\cref{sec:detecting}) identifies a documentation deficiency and writes an entry to the tracker. The entry records the source mechanism, the originating ticket (if any), the gap type, a natural-language description of what is missing, and a suggested fix direction.
\item \textbf{Prioritization.} The system assigns a priority (high, medium, low) based on two signals: the number of distinct tickets in which the gap has surfaced (frequency count), and whether the gap caused an incorrect agent response. Gaps that recur across multiple tickets are prioritized higher.
\item \textbf{Remediation.} The Documentator agent (\cref{sec:fixing}) drains the tracker in priority order, drafting a fix for each gap. The fix passes through self-review and is submitted as a code review for human approval.
\item \textbf{Closure.} Once the human reviewer approves and merges the fix, the gap status transitions to \textit{fixed}. Figure~\ref{fig:cumulative-gaps} shows the cumulative progression from discovery through closure over the deployment period.
\end{enumerate}

\paragraph{Deduplication and recurrence.} When a detection mechanism identifies a gap that matches an existing tracker entry, no duplicate is created. Instead, the system increments the entry's frequency count and appends the new ticket to its list of originating tickets. In our deployment, 11 gaps recurred across multiple tickets (frequency 2--4), and these recurring gaps were disproportionately classified as high priority---confirming that recurrence is a useful prioritization signal.

\paragraph{Discovery sources.} Over the pilot, discovery was dominated by reactive harvesting of resolved tickets (34) and live triage (28), with smaller contributions from the Student's on-demand probing (6) and the Feedback Loop (2); 21 gaps have no recorded detection source (a tracking-metadata omission) and one was logged manually. The proactive Code-Merge Watcher recorded none in-window (\cref{sec:operational}). The tracker's priority breakdown reflects the dominance of \gapmissing{}: because absent documentation affects every user who encounters the topic, these gaps tend to surface repeatedly and escalate in priority.

\section{Code-to-Documentation Registry}
\label{app:registry}

The registry is a version-controlled JSON file that maps each documentation target to the code packages implementing what it describes. Each entry records the package, the role it plays for that document, and the reason the mapping exists. During the pilot it held \metricRegistryTargets{}~documentation targets, and a companion watchlist named \metricWatchedPackages{}~code packages from the service's build pipeline. The Code-Merge Watcher (\cref{sec:detecting}) reads the registry in reverse. For each change merged into a watched package, it looks up the documents mapped to that package and asks an LLM judge whether the change makes any of them stale. A merged change in a package with no registry entry yields no candidate documents, which bounds the watcher's coverage (\cref{sec:limitations}). The Documentator does not yet use the registry; fetching the implementing code at draft time is a planned extension.

\section{Evaluation Protocol Details}
\label{app:eval_details}

This appendix details the protocol shared by the automated analyses---factual correctness (\cref{sec:factual}), usefulness (\cref{sec:pairwise}), and self-improvement (\cref{sec:selfimprove}); the deployment-scale metrics (\cref{sec:operational}) are descriptive. Factual correctness asks whether individual claims in the agent's responses are true; usefulness asks whether the overall response is competitive with the human's; self-improvement asks whether repairing a gap improves the later answer. All three use the judge panel described below.

\subsection{Dataset Construction}
\label{app:dataset}

The evaluation dataset is drawn from a production support queue.
Of the \metricTickets{}~tickets TriAgent triaged during the pilot (its deployment scale), 333 received both an agent response and a recorded human resolution to the same ticket; the remaining triaged tickets lacked a comparable human answer to grade against and fall outside the scored set.
We begin from these 333 pairs, each containing the ticket (title and description), the agent's documentation-grounded response, and the human responder's resolution.

We apply two filters. First, we exclude \emph{mechanical} tickets---requests that require only a rote action (e.g., ``please clear caches for dataset X'') with no diagnostic reasoning. Second, we apply minimum length thresholds (agent response $\geq$50 characters, human response $\geq$20 characters) to remove cases where either party gave a trivially short acknowledgment rather than a substantive answer.

After filtering, we further exclude seven non-answers---tickets the agent correctly declined because a human had already resolved the thread or it was internal-engineering discussion, leaving nothing to grade. This leaves \textbf{192 answered pairs} (Table~\ref{tab:inclusion-flow}), the set used throughout, spanning live triage (agent responded in real time) and retrospective re-triage of historically resolved tickets; the human responses come from the original resolution, written without knowledge that an agent comparison would later occur.

\begin{table}[h]
\centering
\small
\begin{tabular}{lr}
\toprule
\textbf{Stage} & \textbf{Count} \\
\midrule
Tickets triaged (deployment scale) & \metricTickets{} \\
\midrule
Pairs with both agent \& human response & 333 \\
\quad after mechanical + length filters & 199 \\
\quad $-$ non-answers (agent correctly declined) & $-7$ \\
\textbf{Answered pairs (evaluation set)} & \textbf{\metricCPairs{}} \\
\bottomrule
\end{tabular}
\caption{Population inclusion flow. The \metricTickets{}~triaged tickets measure deployment scale and are a \emph{distinct} population; the scored-evaluation set is built separately from the 333~pairs that received both an agent and a human response (live and retrospective), filtered and stripped of seven non-answers to the \metricCPairs{}~answered pairs.}
\label{tab:inclusion-flow}
\end{table}

\paragraph{Confidence intervals.}\label{app:ci} All reported proportions carry Wilson score 95\% confidence intervals. For $x$ successes in $n$ trials with $z=1.96$ and $\hat p = x/n$, the interval is
\[
\frac{1}{1+z^2/n}\left(\hat p + \frac{z^2}{2n} \;\pm\; z\sqrt{\frac{\hat p(1-\hat p)}{n} + \frac{z^2}{4n^2}}\right).
\]
We use the Wilson interval rather than the normal (Wald) approximation because several of our proportions rest on small samples---most notably the \metricDN{}~before/after pairs---where the Wald interval is unreliable and can fall outside $[0,1]$.

\paragraph{Per-class agreement.} Three-judge unanimity varies by class: \metricBUnanCorrect{}\% of assertions the panel labels \textit{correct} and \metricBUnanUnver{}\% of \textit{unverifiable} ones are unanimous, but only \metricBUnanIncorrect{}\% of \textit{incorrect} ones---the judges disagree most about what counts as a contradiction. This is why agreement on the verifiable subset ($\kappa=\metricBKappaVer{}$) is lower than the overall $\kappa=\metricBKappa{}$.

\subsection{Judge Panel}
\label{app:judges}

All automated evaluation uses a panel of three judges (\emph{Models~B, C, D}) drawn from three \emph{different model families}, none of them the family of the agent's generation model (\emph{Model~A}); all model names are anonymized. Each item is scored by all three at temperature~0 (for reproducibility), and we report the majority verdict with inter-judge agreement (Fleiss' $\kappa$). We call two models the same \emph{family} when they come from the same developer. A fourth judge (\emph{Model~E}) from the same family as Model~A is run only as a probe. Prior work documents self- and family-preference in LLM judges \citep{spiliopoulou2025playfavorites,zheng2023mtbench}, so we keep Model~E out of every headline number. Table~\ref{tab:probe} compares its verdicts with the panel's. The probe does not consistently favor the agent. On correctness it is more decisive in both directions. On usefulness it prefers the agent less often than the panel does. On before/after it prefers the post-repair answer more often.

\begin{table}[h]
\centering
\small
\begin{tabular}{lrr}
\toprule
 & Panel & Probe \\
\midrule
\multicolumn{3}{l}{\emph{Correctness} (\metricBAssertions{} assertions)} \\
\quad Correct & \metricBCorrect{} & \metricProbeBCorrect{} \\
\quad Incorrect & \metricBIncorrect{} & \metricProbeBIncorrect{} \\
\quad Unverifiable & \metricBUnverifiable{} & \metricProbeBUnverifiable{} \\
\quad Correct by this judge only & \metricProbeBLeanPanel{} & \metricProbeBLeanCorrect{} \\
\multicolumn{3}{l}{\emph{Usefulness} (\metricCPairs{} pairs)} \\
\quad Agent preferred by this judge only & \metricProbeCLeanPanel{} & \metricProbeCLeanAgent{} \\
\multicolumn{3}{l}{\emph{Before/after} (\metricDN{} pairs)} \\
\quad Post-repair answer preferred & \metricDAfterN{} & \metricProbeDAfter{} \\
\bottomrule
\end{tabular}
\caption{Same-family probe (Model~E) against the cross-family panel majority. The panel has no majority on \metricBTie{} assertions.}
\label{tab:probe}
\end{table}

\subsection{Factual Correctness}

The factual correctness evaluation operates at assertion granularity through a two-pass pipeline:

\paragraph{Pass 1 --- Assertion extraction.} The judge receives the agent's full response and extracts every discrete factual claim as a short, self-contained statement. Claims include causal attributions, specific values or configurations, references to documentation, diagnostic conclusions, and procedural recommendations. Greetings, opinions, hedged language, and questions back to the user are excluded. The prompt template is provided in \cref{app:prompts}.

\paragraph{Pass 2 --- Assertion verification.} Each extracted assertion is verified by the panel against the human's response as the single authoritative reference, under a \emph{mutually-exclusive} rule:
\begin{itemize}

\item \textit{incorrect} --- the reference addresses the claim and contradicts it;
\item \textit{correct} --- the reference addresses the claim and is consistent with it;
\item \textit{unverifiable} --- the reference does not address the claim.
\end{itemize}

Silence is \textit{unverifiable}, never \textit{correct}: a claim is labeled \textit{correct} only when the reference actually speaks to it and agrees. This closes an overlap in the earlier single-judge version, where ``not contradicted'' and ``not addressed'' could both apply to the same claim, and it is why the unverifiable share is large---many terse human resolutions do not address the agent's specific claims. We report accuracy over \emph{verifiable} assertions \emph{and} over \emph{all} assertions, both labeled, rather than a single headline that hides the denominator. The evaluation runs over the 192 answered pairs.

\subsection{Usefulness (Pairwise Preference)}

Each of the 192 pairs is presented to the panel in blinded A/B format---the ticket, then ``Response~A'' and ``Response~B''---with the agent's response assigned to A or B at random per ticket (by a hash of the ticket id, so the assignment is fixed and reproducible). The judges score three dimensions:
\begin{itemize}

\item \textbf{Actionability} --- which response better equips the reader to take the correct next action?
\item \textbf{Completeness} --- which covers more relevant diagnostic information?
\item \textbf{Overall preference} --- which would you prefer to send as the official answer?
\end{itemize}

For each dimension a judge chooses ``A~better'', ``B~better'', or ``tie''; the reported verdict is the cross-family majority on overall preference, with Fleiss' $\kappa$.

\paragraph{Length control.} Because pairwise preference tracks response length, we report the length-comparable subset (the two responses within a $2\times$ length ratio) alongside the full set. A separate coverage pass asks, for each pair, whether the agent raised a correct material point the human omitted, and whether the human raised one the agent omitted.

\subsection{Self-Improvement (Before/After)}

Confidence is the triage's own self-reported label; ``not high-confidence'' ($=$\textsc{medium}/\textsc{low}) covers \metricDNotHighConf{} of the \metricDMatched{}~matched before/after pairs (the rest \textsc{high}, with little room to improve; \metricDNoConf{} unlabeled), of which \metricDNoBefore{}~lack a stored earlier answer---leaving the \metricDN{}~pairs we compare against a later re-answer produced under the updated documentation. The comparison is blinded: the panel sees the two answers as A/B without being told which came first, and reports which better resolves the ticket. We corroborate causal attribution against the documentation repository's revision history---for the gaps the loop tagged, the commit that updates the answering documentation postdates the gap's discovery---rather than relying on the tracker's ``fixed'' flag.

\section{Self-Improvement Case Studies}
\label{app:case_study}

We illustrate the self-improvement loop with two examples from deployment---demonstrations of the loop operating end to end, chosen for clarity rather than as a representative sample. They show both the simplest case (a single knowledge base addition suffices) and the full end-to-end pipeline (gap surfaced from customer interaction, fix drafted autonomously, merged via code review).

\subsection{Anatomy of a Triage Response}
Every triage produces a structured record whose fields feed the gap-detection loop directly. A stylized example (the ramp-schedule scenario of \cref{sec:intro}):
\begin{quote}\small
\textbf{Ticket.} ``Why does the dashboard show a launch date earlier than when I created the experiment?''
\begin{itemize}
\item \emph{response:} ``The date shown is when traffic allocation began under the ramp schedule, not when the experiment was created\dots''
\item \emph{confidence:} \textsc{high}
\item \emph{sources\_cited:} [ramp-schedule guide \S2; experiment-lifecycle overview]
\item \emph{gap\_found:} true \quad \emph{gap\_type:} \gapambiguous{} \quad \emph{target\_doc:} experiment-lifecycle overview
\item \emph{suggested\_fix:} clarify that the displayed ``launch date'' refers to ramp start
\end{itemize}
\end{quote}
The \emph{confidence} field gates the self-improvement selection (\cref{sec:selfimprove}); \emph{gap\_found}, \emph{gap\_type}, \emph{target\_doc}, and \emph{suggested\_fix} feed the Documentator and the improvement tracker.

\subsection{Case 1: Knowledge Base Addition}

\paragraph{Initial state.} A user asks why the analysis engine's variance estimates are conservative relative to an alternative estimator. The agent retrieves all available documentation but finds no authoritative source addressing this specific comparison. It responds with low confidence:

\begin{quote}
\small\textit{``I cannot confirm from available documentation why the variance estimates differ between the two systems. Based on what I can find, both use standard estimators, but I do not have a source that explains the conservatism you are observing. Flagging this for expert input.''}
\end{quote}

The structured learning step classifies this as a documentation gap of type \gapmissing{}: the topic has no knowledge base coverage.

\paragraph{Documentation fix.} A technical paper explaining the finite-population regression variance bound---and why the engine's estimates are conservative by construction---is added to the knowledge base. No agent code or prompt changes are made.

\paragraph{Post-fix state.} On the next occurrence of a related question, the agent retrieves the newly added paper, cites it directly, and provides a definitive answer with high confidence:

\begin{quote}
\small\textit{``The variance estimates are conservative by design. The engine uses a finite-population regression variance bound [citing specific paper, \S3.2] which is provably conservative relative to the Neyman estimator used by the alternative system. This is not a bug---it reflects a deliberate choice to avoid anti-conservative inference in finite samples.''}
\end{quote}

The self-improvement step reports: ``No documentation gap found.''

\paragraph{Observation.} The entire loop---from gap detection to resolution---required only a knowledge base update. No retraining, no prompt modification, and no code change. The agent's behavior improved solely because its retrieval source improved.

\subsection{Case 2: Full Pipeline (Ticket to Merged Fix)}

This case illustrates the complete system operating end-to-end, including a subtlety: the agent initially fails to recognize a gap, and customer feedback is needed to surface it.

\paragraph{Ticket.} A user asks whether a foundational technical document is outdated because it does not mention a recently introduced methodology. The user has independently found two related documents and wants to understand how the three relate.

\paragraph{Initial triage.} The agent correctly explains the relationship between the three documents---the foundational document remains authoritative, and the newer documents extend (rather than supersede) it. Confidence is high. The structured learning step reports ``No documentation gap found,'' reasoning that the information exists across the three documents and the separation is an organizational choice.

\paragraph{Customer feedback.} The user replies suggesting that a cross-reference note in the foundational document would prevent future users from going on ``the same scavenger hunt.'' This reframes the issue: the gap is not missing content but missing navigational structure between related documents.

\paragraph{Gap detection on re-triage.} On re-triage with the new correspondence, the agent recognizes the customer's confusion as the gap signal itself. It writes a structured entry to the improvement tracker: type \gapambiguous{}, target document identified, suggested fix is a cross-reference paragraph.

\paragraph{Documentator execution.} The Documentator agent reads the target document, drafts a ``Related documents'' paragraph with proper citations to both companion papers, and submits the fix as a code review.

\paragraph{Closure.} The human reviewer approves the change. The foundational document now cross-references its companion papers, preventing the navigational confusion that generated the original ticket.

\subsection{Discussion}

These two cases illustrate complementary aspects of the self-improvement loop:

\begin{itemize}

\item \textbf{Case 1} demonstrates that the loop requires no retraining---only knowledge base expansion. Most gaps are of this kind (missing content): the system identifies what is missing, a human adds the documentation, and the agent benefits on subsequent queries.
\item \textbf{Case 2} demonstrates the system's ability to learn from customer feedback as well as from its own failures. The agent's initial triage was correct, yet the customer's response revealed a usability gap invisible to a purely content-focused analysis. It also shows the full autonomous pipeline: from gap detection through drafted fix to merged code review, with the human's role shifted from author to reviewer.
\end{itemize}

In both cases, each resolved ticket strengthens the documentation that informs future responses---the implicitly-supervised repair loop described in \cref{sec:conclusion}.

\section{Judge Prompt Templates}
\label{app:prompts}

We provide the exact prompt templates used in our LLM-as-judge evaluation for reproducibility. All evaluations use the judge model (\emph{Model~B}) at temperature~0. Each prompt is shown as delivered to the model, with placeholders (in braces) filled at runtime from the evaluation pair.

\subsection{Factual Correctness: Assertion Extraction}

This prompt is applied once per agent response. It receives the full response text and returns a list of discrete factual claims. The key design choice is the explicit exclusion list (greetings, opinions, hedges, questions)---without it, the judge tends to extract subjective statements that cannot be meaningfully verified.

\begin{quote}
\small
\textit{You are extracting factual claims from a technical support response. List every concrete factual assertion the response makes---things that are either true or false, verifiable against documentation or system behavior.}

\textit{Extract each factual claim as a short, self-contained statement. Include:}
\begin{itemize}

\small
\item \textit{Causal claims (``X happens because Y'')}
\item \textit{Specific values, thresholds, or configurations mentioned}
\item \textit{References to specific documentation, tools, or system behavior}
\item \textit{Diagnostic claims (``the root cause is X'')}
\item \textit{Procedural claims (``to fix this, do X then Y'')}
\end{itemize}

\textit{Do NOT extract: greetings or politeness, opinions or subjective assessments, questions asked back to the customer, hedging language (``this might be'', ``possibly'').}

\textit{Output format: respond with ONLY a JSON array of strings (no markdown, no commentary):}

\textit{[``assertion 1'', ``assertion 2'', ``assertion 3'', ...]}

\textit{If the response contains no verifiable factual claims, return: []}
\end{quote}

\subsection{Factual Correctness: Assertion Verification}

This prompt is applied once per extracted assertion. It receives the assertion, the original ticket context, and the human's response (treated as reference answer). The three labels are mutually exclusive: \textit{correct} requires the reference to actually address the claim and agree with it; \textit{incorrect} requires a direct factual conflict; \textit{unverifiable} is used when the reference is silent on the claim. Silence is never \textit{correct}.

\begin{quote}
\small
\textit{You are verifying whether a factual claim made in a support response is correct, given a reference answer from a domain expert.}

\textit{The claim: ``\{assertion\}''}

\textit{Context --- the customer's question:}

\textit{Title: \{ticket\_title\}}

\textit{\{ticket\_description\}}

\textit{Reference answer (from the domain expert who resolved this ticket):}

\textit{\{ic\_response\}}

\textit{Determine whether the claim is:}
\begin{itemize}

\small
\item \textit{``incorrect'' --- the reference addresses the claim and contradicts it}
\item \textit{``correct'' --- the reference addresses the claim and is consistent with it}
\item \textit{``unverifiable'' --- the reference does not address the claim}
\end{itemize}

\textit{Important: the three labels are mutually exclusive. ``correct'' requires the reference to actually speak to the claim and agree; if the reference is silent on the point, the label is ``unverifiable'', never ``correct''.}

\textit{Output format: respond with ONLY a JSON object (no markdown, no commentary):}

\textit{\{``verdict'': ``correct$|$incorrect$|$unverifiable'', ``rationale'': ``1 sentence explanation''\}}
\end{quote}

\subsection{Pairwise Preference}

This prompt is applied once per evaluation pair. It receives the ticket and two blinded responses (A and B, with assignment determined by the counterbalancing configuration). The key design choices are: (a)~three independent sub-questions that decompose preference into actionability, completeness, and overall judgment; (b)~forced ternary (A/B/tie) with no numeric scoring; and (c)~a mandatory rationale per dimension to encourage deliberation before the verdict.

\begin{quote}
\small
\textit{You are evaluating two responses to a technical support ticket. Both responses attempt to help the customer resolve their issue. Your job is to compare them on three specific criteria.}

\textit{The support ticket:}

\textit{Title: \{ticket\_title\}}

\textit{\{ticket\_description\}}

\textit{Response A: \{response\_a\}}

\textit{Response B: \{response\_b\}}

\textit{Answer each question independently. For each, choose exactly one of: ``A'' (Response A is better), ``B'' (Response B is better), or ``tie'' (both are roughly equivalent on this criterion).}

\textit{Q1---Actionability: if you were the on-call engineer receiving one of these responses as guidance, which would better equip you to take the correct next action without further research?}

\textit{Q2---Completeness: which response covers more of the relevant diagnostic information for this specific ticket? Consider: root cause identification, relevant context, and next steps.}

\textit{Q3---Overall preference: setting aside formatting and style, which response would you prefer to send to the customer as the official answer to their ticket?}

\textit{Output format: respond with ONLY a JSON object (no markdown, no commentary):}

\textit{\{``actionability'': \{``choice'': ``A$|$B$|$tie'', ``rationale'': ``1 sentence why''\}, ``completeness'': \{...\}, ``overall'': \{...\}\}}
\end{quote}

\end{document}